\documentclass[osajnl,showpacs]{revtex4}  %% REVTeX 4.0
\begin{document}

\title{Optical coherence of planar microcavity emission.}

\author{R. F. Oulton, J. W. Gray, P. N. Stavrinou and G. Parry.}

\affiliation{Centre for Electronic Material and Devices, \\ Imperial College of Science, Technology and Medicine, London.}

\begin{abstract}

An analytical expression for the self coherence function of a microcavity and partially coherent source is derived from first principles in terms of the component self coherence functions. Excellent agreement between the model and experimental measurements of two Resonant Cavity LEDs (RCLEDs) is evident. The variation of coherence length as a function of numerical aperture is also described by the model. This is explained by a microcavity's angular sensitivity in filtering out statistical fluctuations of the underlying light source.  It is further demonstrated that the variable coherence properties of planar microcavities can be designed by controlling the underlying coherences of microcavity and emitter whereby coherence lengths ranging over nearly an order of magnitude could be achieved.

\end{abstract}


\maketitle

The last two decades have seen widespread use of optical microcavities, both for experimental physics and commercial applications.  Microcavities redistribute emission from an underlying source and depending on the ensuing radiation pattern allow light collection for use elsewhere. Commercially available microcavity devices such as Resonant Cavity Light Emitting Diodes (RCLEDs) use the planar microcavity geometry to increase the extraction efficiency of spontaneous emission from materials with high dielectric constants. \cite{Ben98} More recently microcavities have been used to spectrally and spatially isolate quantum dot emitters to increase the efficiency of single photon production. \cite{Mor01,San02}

Recent work on planar microcavities has identified the dependence of Numerical Aperture (NA) on emission properties such as spectral linewidth \cite{Sta99,Gra01,Oul01} and coherence length,\cite{Ric03} the latter of which is the focus of the following paper. Note that, migration of these results to the spectral domain are trivial due to the implicit Fourier relationship with the coherence domain. In addition to the choice of domain, the name {\it coherence} has been used, instead of {\it time}. Coherence highlights the statistical properties of light and not necessarily to the time dependence of the emission process that produces the light. This distinction between the light and the emission event that produced it is necessary for the descriptions presented in this letter.

The coherence length is a relevant attribute of any optical device as it defines the length scales over which mutual interference occurs. Applications such as Low Coherence Interferometry, for non invasive medical imaging \cite{Sch99} (also known as Optical Coherence Tomography ) and Optical Time Domain Reflectometry (OTDR), for ranging measurements in optical components \cite{You87} and surface mapping in integrated circuits \cite{Kin90}, rely on the coherence of source being both small enough to eliminate coherent reflections from distant objects and large enough to examine detail on the relevant length scale. The correlations of low coherence sources are also applicable to schemes of all-optical routing.\cite{Sam90} 

The variable coherence properties of RCLEDs \cite{Ric03} make them ideal as a light source for applications where a range of length scales need to be analysed where previously, multiple light sources would have been required. This letter examines the physics of variable emission coherence from planar microcavities and indicates methods by which this aspect of commercial devices such as RCLEDs could be engineered. Although experimental results of RCLEDs are used in this paper, the results are applicable to planar microcavities in general.

The coherence domain picture of microcavity emission requires the definition of the self coherence properties of a light source. The self coherence function of an emitter, $\Gamma_E(\tau)$, defined in Eqn.~(\ref{eqn0}) is synonymous with the spectral distribution function by the Wienner-Kinchin theorem. \cite{Born&Wolf}
\begin{eqnarray}\label{eqn0} \Gamma_E(\tau) = \int_{-\infty}^{\infty} E(t)E^*(t- \tau)dt \end{eqnarray}

A derivation for $\Gamma(\tau,\theta_c)$, the self coherence function of a combined emitter and microcavity system,  is too involved to be described here from first principles. Instead, a key result from the spectral domain is used in combination with the Wienner-Kinchin theorem. The emission intensity from an emitter microcavity system is given by the spectral overlap of the optical transfer function of the bare cavity, $H(\omega,\theta_c)$, which is a function of the angle relative to the emission region, $\theta_c$, and the field distribution of the underlying source $|E(\omega)|^2$, which is assumed to be isotropic. \cite{Dre74,Dep94}
\begin{eqnarray}\label{eqn8} I(\omega) = H(\omega,\theta)|E(\omega)|^2 \end{eqnarray}

Therefore the coherence function for the microcavity can be written as a convolution expression:
\begin{eqnarray}\label{eqn6} \Gamma(\tau,\theta_c) & = &  \frac{\Gamma_0}{2\pi} \int_{-\infty}^{\infty} \Gamma_E(\tau - t) \Gamma_C(t,\theta_c) dt \end{eqnarray}

$\Gamma_0$ is the maximum microcavity enhancement for an emitter tuned to the cavity resonance at an angle $\theta_t$,
\begin{eqnarray}\label{eqn7} \Gamma_0 = \frac{ T_F(\theta_t) \left( 1+ \sqrt{R_F(\theta_t)} \right)^2}{1 - \sqrt{R_F(\theta_t)R_B(\theta_t)}} \end{eqnarray}

Here $T_F(\theta_c)$ and $R_F(\theta_c)$ are the transmission and reflectivity of the front cavity mirror and $R_B(\theta_c)$ is the reflectivity of the back mirror; all function of $\theta_c$, the internal emission angle.
 
$\Gamma(\tau,\theta_c)$ can be greatly simplified by evaluating the definite integral. To do this, the underlying spectral distribution of emission is assumed to be Lorentzian in nature corresponding to exponential decay in the time domain. The transfer function of the cavity, $H(\omega,\theta_c)$ can also be approximated by Lorentzians in the immediate vicinity of the cavity modes, $\omega_c(\theta_c)$ as the free spectral range of longitudinal modes in microcavities is usually much greater than the spectral width of the source.
\begin{eqnarray}\label{eqn9} \Gamma(\tau,\theta_c) = \frac{\Gamma_0}{2\pi} \int_{-\infty}^{\infty} \exp{(-i\omega_E (\tau - t) - |\tau - t|/\tau_E)} \exp{(-i\omega_c(\theta_c)t - |t|/\tau_c(\theta_c))} dt \end{eqnarray}

Here $\omega_E$ and $\omega_c(\theta_c)$ are the frequencies of the emission and cavity resonances respectively and $\tau_E$ and $\tau_c(\theta_c)$ are the coherence times of the underlying source and cavity resonances respectively. Note that $\tau_E$ and $\omega_E$ are assumed to be constant with emission angle. Since $\Gamma(\tau,\theta)$ is not necessarily symmetric, consider the expansion of the integral in Eqn.~(\ref{eqn9}) for $\tau > 0$:
\begin{eqnarray}\label{eqn10} \Gamma(\tau,\theta_c) & = &\frac{\Gamma_0}{2\pi} \int_{\tau}^{\infty} \exp{(-i\Delta \omega(\theta_c) t - K(\theta_c)t)} \exp{(-i\omega_E\tau + \tau/\tau_E)} dt \nonumber \\ & & + \frac{\Gamma_0}{2\pi} \int_{0}^{\tau} \exp{(-i\Delta \omega(\theta_c) t - K^\prime(\theta_c)t)} \exp{(-i\omega_E\tau - \tau/\tau_E)} dt \nonumber \\ & & + \frac{\Gamma_0}{2\pi} \int_{-\infty}^{0} \exp{(-i\Delta \omega(\theta_c) t + K(\theta_c)t)} \exp{(-i\omega_E\tau - \tau/\tau_E)} dt  \end{eqnarray}

Here, $K(\theta_c) = 1/\tau_c(\theta_c) + 1/\tau_E$ and $K^\prime(\theta_c) = 1/\tau_c(\theta_c) - 1/\tau_E$. Evaluating the integrals of Eqn.~(\ref{eqn10}) gives the self coherence function for the microcavity at an internal angle $\theta_c$,
\begin{eqnarray}\label{eqn11} \Gamma(\tau,\theta_c) =\frac{\Gamma_0}{2\pi} \left\{ \frac{\Gamma_C(\tau,\theta_c)}{i\Delta \omega(\theta_c) + K(\theta_c)} - \frac{\Gamma_C(\tau,\theta_c)}{i\Delta \omega(\theta_c) + K^\prime(\theta_c)} + \frac{\Gamma_E(\tau)}{i\Delta \omega(\theta_c) + K^\prime(\theta_c)} -  \frac{\Gamma_E(\tau)}{i\Delta \omega(\theta_c) - K(\theta)} \right\}  \end{eqnarray}

The real and imaginary parts of this expression are evaluated to be:
\begin{eqnarray}\label{eqn12} \Gamma(\tau,\theta_c) & = & \frac{\Gamma_0}{2\pi} \left\{ \frac{K^\prime(\theta_c)}{\Delta \omega(\theta_c)^2 + K^{\prime2}(\theta_c)} \left( \Gamma_E(\tau) - \Gamma_C(\tau,\theta_c) \right) +  \frac{K(\theta_c)}{\Delta \omega(\theta_c)^2 + K^{2}(\theta_c)} \left(\Gamma_E(\tau) + \Gamma_C(\tau,\theta_c) \right) \right. \nonumber \\ & & - \left. i \Delta \omega (\theta_c) \left[ \left(\frac{1}{\Delta \omega(\theta_c)^2 + K^{\prime2}(\theta_c)} - \frac{1}{\Delta \omega(\theta_c)^2 + K^{2}(\theta_c)}  \right)  \left(\Gamma_E(\tau) - \Gamma_C(\tau,\theta_c) \right)  \right] \right\} \end{eqnarray}

The expansion of the integral in Eqn.~(\ref{eqn9}) for $\tau < 0$ gives $\Gamma(\tau<0,\theta_c) = \Gamma^*(\tau>0,\theta_c)$. Such a symmetry is necessary in this case so that the Fourier transform gives the real-valued spectral density function for cavity emission. Examination of Eqns.~(\ref{eqn11}) and~(\ref{eqn12}) highlight the correct asymptotic response: If $\tau_c(\theta_c) >> \tau_E$ then $\Gamma(\tau,\theta_c = 0, \Delta \omega(\theta_c) = 0) \propto \Gamma_C(\tau,\theta_c)$, the coherence function of the cavity. For this condition $K = - K^\prime \approx 1/\tau_E$ such that:
\begin{eqnarray}\label{eqn13} \Gamma(\tau,\theta_c) = \frac{\Gamma_0}{\pi}\frac{K}{\Delta \omega(\theta_c)^2 + K^2} \Gamma_C(\tau,\theta_c)  \end{eqnarray}

Similarly, if $\tau_E >> \tau_c(\theta_c)$ then $\Gamma(\tau,\theta_c = 0, \Delta \omega(\theta_c) = 0) \propto \Gamma_E(\tau)$, the coherence function for the underlying emission.  For this condition $K = K^\prime \approx 1/\tau_c(\theta_c)$ such that:
\begin{eqnarray}\label{eqn14} \Gamma(\tau,\theta_c) =  \frac{\Gamma_0}{\pi}\frac{K(\theta_c)}{\Delta \omega(\theta_c)^2 + K(\theta_c)^2} \Gamma_E(\tau) \end{eqnarray}

These two limiting cases are also of particular interest to commercial applications. For $\tau_c(\theta_c) >> \tau_E$ is typically required for lasing applications, Here, notice that the self coherence function, $\Gamma(\tau,\theta_c) \propto \Gamma_C(\tau,\theta_c)$ is a function of angle, however is solely dependent on the cavity. Engineering the cavity allows direct control of the lasing mode. The second limit, $\tau_E >> \tau_c(\theta_c)$, has only recently become applicable in semiconductor devices with the advent of devices that have isolated single quantum dot light sources. \cite{Mic00,Yua02} The spectrally pure emission from single quantum dot sources within a microcavity will have coherence properties that vary weakly with angle since $\Gamma(\tau,\theta_c) \propto \Gamma_E(\tau)$. The control over emission direction is still available by engineering the cavity's tuning. 

For RCLEDs $\tau_c(\theta_c) > \tau_E$, however, the difference is not large enough to reduce the dynamics to one of the cases described above by Eqns.~(\ref{eqn13}) and~(\ref{eqn14}). The coherence model has been compared with experimental results for tuned and detuned RCLEDs operating near $650$ nm, the details of which can be found in Refs. \onlinecite{Gra01} and \onlinecite{Oul01}. Angle resolved spectra were sampled through a pin hole subtending a half angle $<0.2^\circ$ at the device using a monochromator with a resolution $<1$ nm. These were converted to the coherence domain by inverse Fourier transform. The coherence length, $L_c(\theta_a)$, measured with respect to the angle, $\theta_a$ in air, is defined in Eqn.~(\ref{eqn15}) is used to analyse the correspondence of spectral domain measurements and the coherence model developed here.
\begin{eqnarray}\label{eqn15} L^2_c(\theta_a) =  c^2\frac{\int_{-\infty}^\infty \tau^2 | \Gamma(\tau,\theta_c) |^2 d\tau}{\int_{-\infty}^\infty | \Gamma(\tau,\theta_c) |^2 d\tau} \end{eqnarray}

The fitting parameters for the coherence model are the coherence times of the cavity mode and underlying source, $\tau_c$ and $\tau_E$ respectively and the relative detuning, $\Delta$. Here, it is assumed that the underlying cavity coherence is approximately constant with angle, a fair approximation for high index devices. Figure~\ref{fig1} shows the variation of coherence length, $L_c(\theta_a)$ for the tuned (Fig.~\ref{fig1}. (a)) and detuned (Fig.~\ref{fig1}. (b)) samples. The broken lines represent the limiting case of $\tau_c >> \tau_E$ which is clearly an insufficient description for both RCLED samples. The solid line is a least squares fit of the model and the three unknown model parameters. The correspondence of experiment and theory is excellent for the lower angles. At large angles, it should be noted that optical loss from the doped regions of the RCLEDs may cause the deviation from the theoretical trend. This is supported by the observation that the deviations occur at the same spectral position, i.e. a relative shift in the deviation's angular position of $15^\circ$ due to the relative device tunings. Table~\ref{tab1} shows the fitting parameter values for the RCLED samples in terms of coherence time (and spectral wavelength). This has been done as it is more natural to specify the detuning in terms of wavelength.

\begin{table}[h]
\caption{Table of fitting parameter values given for both coherence (time) and spectral domain.}\label{tab1}
\begin{tabular}{l|rr}
Parameter &  Tuned  & Detuned   \\ \hline
$\tau_c$ & $0.115$ ps ($3.89$ nm) & $0.111$ ps  ($4.04$ nm)  \\ 
$\tau_E$ & $0.034$ ps  ($13.23$ nm) & $0.035$ ps  ($12.77$ nm)  \\ 
$\Delta$ &    ($+0.58$ nm) &    ($-5.50$ nm) \\ \hline
\end{tabular}
\end{table}

Good agreement is apparent between parameters of the tuned and detuned RCLEDs. In addition, the values for the detuning closely match the design values of $0$ nm and $-6$ nm for tuned and detuned devices respectively. The trend itself is also of significant importance; clearly interference between the cavity mode and underlying emission only resolves to a peak in coherence at the tuning angle. This is only visible in the case of the detuned RCLED at approximately $26^\circ$, corresponding to a detuning of $-5.5$ nm.

In the previous study, the density of emission states within the numerical aperture of emission was taken into account. Here, the density of off-axis states increases as $\sin \theta_a$. In addition, the differential change in solid angle with respect to air $d\Omega_a = \sin\theta_a d\theta_a d\phi$ and the cavity $d\Omega_c = \sin \theta_c d\theta_c d\phi$ must be taken into account. By integrating Eqn.~(\ref{eqn12}) over solid angle, the variation of the self coherence function as a function of NA is given by: 
\begin{eqnarray}\label{eqn16} \Gamma(\tau,NA=\sin \theta_a) & = & \int_0^{2\pi}\!\!\!\int_0^{\theta_a} \Gamma(\tau,\theta_c) \frac{d\Omega_c}{d\Omega_a} d \Omega_a \nonumber \\ & = & 2\pi\int_0^{\theta_a} \Gamma(\tau,\theta_c) \frac{\cos \theta_a}{\cos \theta_c} \sin \theta_a d \theta_a \end{eqnarray}

Figure~\ref{fig2}. shows the variation of coherence length as a function of NA for the two RCLEDs under investigation evaluated using the coherence model with the parameters shown in Tab.~\ref{tab1}. Although experimental results for these trends are not available, they do follow the generic trend observed recently in Ref. \onlinecite{Ric03}. The square markers show the coherence length extrema calculated from measured spectra at normal incidence and $NA = 1$ in an integrating sphere offering further confirmation of the result.

The coherence properties of planar microcavities are modelled well by the coherence formula of Eqn.~(\ref{eqn12}). Consider therefore, using this formula to predict how large a range of coherence length variation can be engineered over useful numerical apertures. The reader will notice, that a low NAs, when only a few of the transverse cavity modes are sampled, the coherence is near to that of the cavity. Statistical fluctuations within the cavity are small since the cavity samples the emission source over a long time than its coherence time resulting in a sharper spectral line and enhanced coherence properties. The behaviour is also seen for all other emission angles into air as the cavity lifetime is approximately constant, however, the emission frequency does change with angle. At large NAs, each cavity mode still samples the underlying emission over a longer time scale than the coherence time of the source, however, the sampled emission is reconstituted spectrally, reproducing to some extent the underlying statistical fluctuations. This appears to be a fair description given the observations. Therefore, in order to create a larger range of coherence variation in these device, the finesse of the cavity must be increased and the coherence of the emission source must be minimised. Most relevant, however, will be to increase the cavity finesse, over which most control can be achieved.

To test this concept, consider varying the cavity finesse in the model for the tuned device discussed above. Fig.~\ref{fig3} shows the maximum and minimum coherence lengths for useful NAs for the RCLED device as a function of finesse and for a range of device tunings. Here, {\it useful} NA must be stressed: The maximum coherence corresponds to the NA through which at least $10$ \% of the device power emitted into air is sampled while the minimum coherence is limited by the limitation of coupling optics set here at an angle of $60^\circ$. These limits are illustrated at the top left corner of Fig.~\ref{fig3}.

Despite the restrictions on useful NA, a large variation in coherence length is apparent in this extrapolated example as is evident from Fig.~\ref{fig3}. Increasing the finesse by an order of magnitude, increases the maximum coherence length by just over a factor of $4$. The minimum NA is increased by nearly a factor of $2$. The range of coherence lengths, on the other hand goes from $10 - 20\, \mu$m to $20 - 80\, \mu$m such that it almost spans an entire order of magnitude.

It was evident in the result of Figs.~\ref{fig1} a) and~\ref{fig1} b) that the maximum coherence length occurs at the tuning angle or wavelength. This behaviour is also seen in the results of Fig.~\ref{fig3} where the maximum coherence length occurs for a tuned sample and indeed gives the greatest variation over useful NAs. Quantitatively, this amounts to a $2$ fold increase in maximum coherence for a tuned device compared with a $6$ nm detuned, which would be used for greater power extraction into large NAs. Clearly, high finesse, tuned microcavity devices provide the greatest coherence variation over useful NAs.

The self coherence function for planar microcavities has been modelled using an analytical formula based on the underlying self coherence functions of the emitter and microcavity. The correspondence of the measured coherence of two RCLED samples and the coherence model described by Eqns.~(\ref{eqn12}) and~(\ref{eqn16}) is compelling. Firstly, a good function fit to the coherence lengths of Fourier transformed spectra sampled at discrete angles was found. Furthermore, the same fitting parameters reproduced closely, the coherence variation as a function of NA, by integrating over the whole set of emission states.\cite{Ric03} 

The central result of this analysis is the observation that the coherence of the device is dependent on the underlying coherence of the cavity and emitter. Indeed the underlying statistical fluctuation of the light source are exposed when varying NA. At low NAs, the coherence length is close to that of the microcavity. At large NAs, the coherence length is closer to that of the underlying emitter. A key aspect of the modelling shows that the coherence properties of planar microcavities can be engineered through microcavity design and selection of an emitter with suitable coherence properties. Here, the model was used to predict a larger range of coherence variation by increasing the finesse of the microcavity. Coherence lengths across an order of magnitude could be achieved if these design consideration were considered carefully.

These observations suggest that variable coherence is a result of a filtering mechanism; this is also the spectral domain picture of microcavity emission. However, in the coherence domain, it is the statistical fluctuations of the light source that are filtered by the slow microcavity sampling time. This is the only robust interpretation that can reconcile the difference in coherence between planar microcavity emission viewed through small and large NAs and raises questions about the relationship between the emission lifetime and coherence time in any type of microcavity device.

\newpage

\section*{List of Figure Captions}

\begin{figure}[h]\centerline{\scalebox{1}{\includegraphics{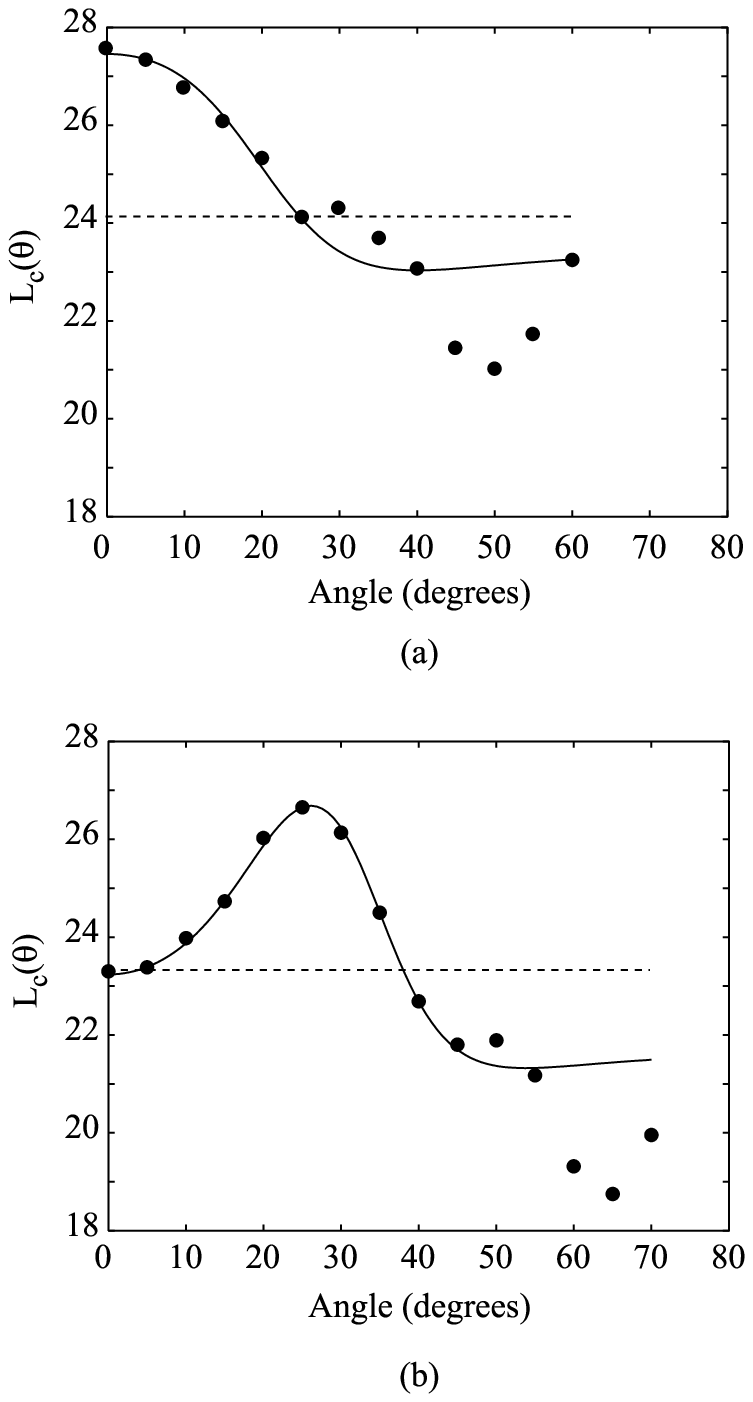}}}
\caption{Comparison of experimental and model results for coherence length variation with observation angle for (a) a tuned RCLED and (b) a detuned RCLED. fig1.eps}\label{fig1}
\end{figure}

\noindent Fig. \ref{fig1}. Comparison of experimental and model results for coherence length variation with observation angle for (a) a tuned RCLED and (b) a detuned RCLED.

\begin{figure}[h]\centerline{\scalebox{1}{\includegraphics{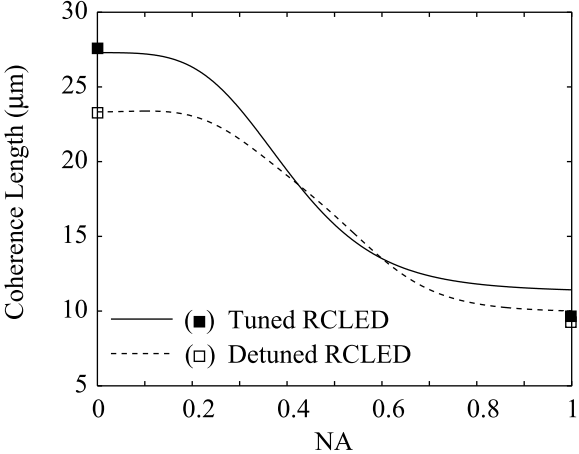}}}
\caption{Coherence length variation as a function of Numerical Aperture generated using empirical model data in the coherence model for tuned and detuned RCLEDs. Markers at NA extrema denote experimental points determine by measurements at low NA and from an integrating sphere. fig2.eps}\label{fig2}
\end{figure}

\noindent Fig. \ref{fig2}. Coherence length variation as a function of Numerical Aperture generated using empirical model data in the coherence model for tuned and detuned RCLEDs. Markers at NA extrema denote experimental points determine by measurements at low NA and from an integrating sphere.

\begin{figure}[h]\centerline{\scalebox{1}{\includegraphics{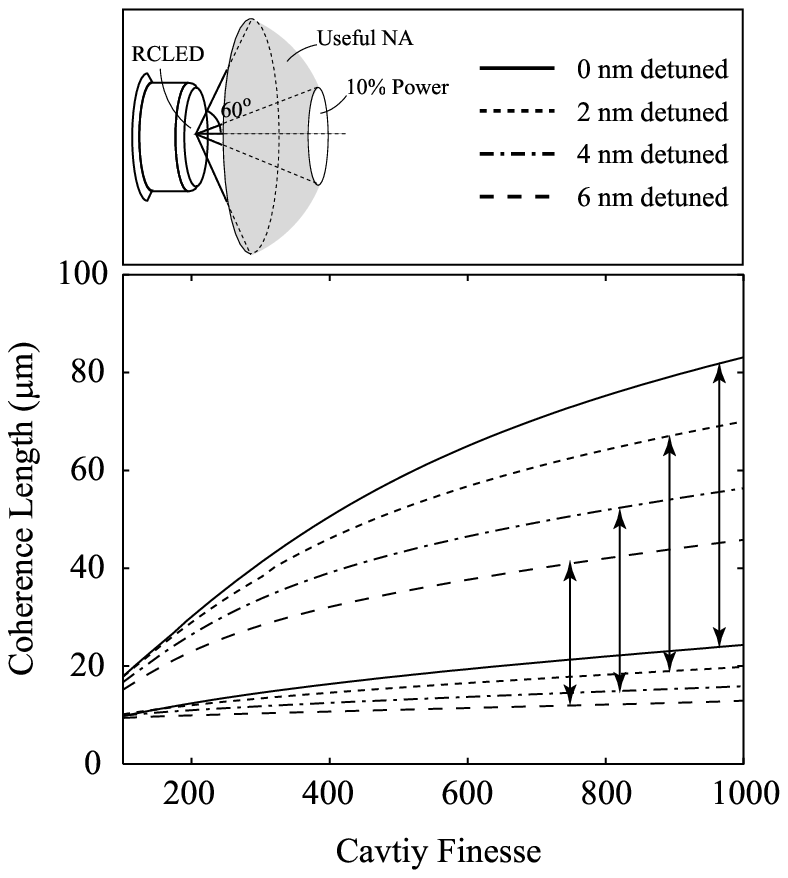}}}
\caption{Predictions of maximum and minimum coherence as a function of microcavity finesse and detuning with respect to the emission source across the useful emission NA (see text). The diagram in the top left corner shows highlights the definition of useful NA. fig3.eps}\label{fig3}
\end{figure}

\noindent Fig. \ref{fig3}. Predictions of maximum and minimum coherence as a function of microcavity finesse and detuning with respect to the emission source across the useful emission NA (see text). The diagram in the top left corner shows highlights the definition of useful NA.

\newpage

\begin{thebibliography}{}

\bibitem{Ben98} H. Benisty et al, IEEE J. Quantum Electron. {\bf34}, 9, 1612-1631 (1998).
\bibitem{Mor01} E. Moreau et al, Appl. Phys. Lett., {\bf 79}, 18, 2865 - 2867 (2001).
\bibitem{San02} C. Santori et al, Nature, {\bf 419}, 10, 594 - 597 (2002).
\bibitem{Sta99} P. N. Stavrinou et al, J. Appl. Phys. {\bf 86}, 6, 3475-3477 (1999).
\bibitem{Gra01} J.W. Gray et al, ``Angular emission profiles and coherence length measurements of highly efficient, low-voltage resonant-cavity light-emitting diodes operating around 650 nm." in \emph{Light-Emitting Diodes: Research, Manufacturing, and Applications V}, H. Walter Yao, E. F. Schubert, eds., Proc. SPIE {\bf4278}, 81-89 (2001).
\bibitem{Oul01} R. F. Oulton et al, Optics Comms. {\bf 195}, 5-6, 327 - 338 (2001).
\bibitem{Ric03} R. C. Coutinho et al, J. Lightwave Technol. {\bf 21}, 1, 149-154 (2003).
\bibitem{Sch99} J.M. Schmitt, IEEE J. Sel. Topics in Quant. Electron. {\bf 5}, 4, 1205-1215 (1999).
\bibitem{You87} R. C. Youngquist et al, Opt. Lett., {\bf 12}, 158 (1987)
\bibitem {Kin90} G. S. Kino et al, Appl. Opt., {\bf 29}, 3775, (1990)
\bibitem{Sam90} D. D. Sampson et al, Electron. Lett. {\bf 26}, 19, 1550 - 1551 (1990).
\bibitem{Born&Wolf} M. Born \& E. Wolf, \emph{Principles of Optics, 6$^{th}$ (Corrected) Edition}, (Cambridge, 1999).
\bibitem{Dre74} K. H. Drexhage, ``Progress in Optics" ed. E. Wolf, Vol. 12, Ch. 6, (1974).
\bibitem{Dep94} H. G. Deppe et al, J. Mod. Opt. {\bf 41}, 2, 325-344 (1994).
\bibitem{Mic00} C. Santori et al, Nature {\bf 419}, 594-597 (2002).
\bibitem{Yua02} Z. Yuan et al, Science {\bf 295}, 102-105 (2002).


\end{thebibliography}
\end{document}